\documentclass[11pt,a4paper]{article}
\usepackage[utf8]{inputenc} 

\usepackage{amsmath,amssymb,amscd,amsfonts,mathabx, textcomp}
\usepackage{amsthm}
\usepackage{cancel}
\usepackage{floatflt}

\begin{document}

\title{\bf Perturbations in $f(\mathbb T)$ cosmology\\ and the spin connection}

\author{Alexey Golovnev\\ \\
\it Centre for Theoretical Physics, The British University in Egypt\\ \it 11837 El Sherouk City, Cairo Governorate, Egypt\\  \small agolovnev@yandex.ru}
\date{}

\maketitle

\abstract{In this short paper we explain how the spin connection variables should be introduced into the perturbation theory of the spatially flat cosmology in $f(\mathbb T)$ gravity with the background tetrad $e^a_{\mu}=a(t)\cdot\delta^a_{\mu}$ and zero background spin connection. In passing, we also correct some small mistakes in our previous work on this subject.}

\section{Introduction}

Among other approaches to modified gravity, teleparallel theories are quickly becoming more and more popular. And since they are actively used for cosmological model building \cite{review}, it is important to have a clear understanding of how the cosmological perturbation theory works in them. This issue has been thoroughly discussed in our previous work \cite{GK} where we have adopted the classical formulation of $f(\mathbb T)$ gravity which makes use of zero spin connection.

The choice of zero spin connection entails violation of local Lorentz symmetry in the space of tetrads, though the model can be easily covariantised by introducing an arbitrary inertial spin connection \cite{Martin, GKS}. Moreover, some authors would even insist that only the covariant version should be used. Then, it seems important to understand how the perturbation theory works in the covariant formulation, too, even though it can be easily shown generically that the theory experiences absolutely no modification beyond the aesthetic one if rewritten in the covariant form \cite{GKS, Manuel}.

Such an attempt has indeed been undertaken in a recent paper \cite{Topor}. However we feel that many things there remain inconclusive. The calculations are done by a brute force approach with unclear results while, in reality, inclusion of the spin connection into perturbation theory is a very simple task which can be understood in a beautiful way. 

In the present paper we would like to explain how we think the perturbation theory must be done. In Section 2 we review the issue of cosmological perturbations in $f(\mathbb T)$ models of gravity and also correct some mistakes from our previous work \cite{GK}. In Section 3 we consider perturbations of a flat spin connection. In Section 4 we present a very simple and clear way of incorporating them into the cosmological perturbation theory. Finally, in Section 5 we conclude.

\section{Review and corrections to the previous work}
The action of $f(\mathbb T)$ gravity is
\begin{equation}
\label{action}
S=-\int d^4 x \| e\|\cdot f\left({\mathbb T}\right)
\end{equation}
which yields the equations of motion
\begin{equation}
\label{eom}
f_{T}\mathop{G_{\mu\nu}}\limits^{(0)}+f_{TT}S_{\nu\mu\alpha}\partial^{\alpha}{\mathbb T}+\frac12 \left(f-f_{T}{\mathbb T}\right)g_{\mu\nu}=8\pi G\cdot\Theta_{\mu\nu}
\end{equation}
where $(0)$ over an object means that it has been calculated in terms of the Levi-Civita connection instead of the teleparallel one.

The notations are as follows. The spacetime connection is given by the tetrad $e$ and the spin connection $\omega$ as
$$\Gamma^{\alpha}_{\mu\nu}= e_A^{\alpha}\left(\partial_{\mu}e^A_{\nu}+\omega^A_{\hphantom{A}\mu B}e^B_{\nu}\right),$$
the torsion tensor is simply $T^{\alpha}_{\hphantom{\alpha}\mu\nu}=\Gamma^{\alpha}_{\mu\nu}-\Gamma^{\alpha}_{\nu\mu}$, the superpotential and the torsion scalar are defined by
\begin{equation}
\label{superpot}
S^{\alpha\mu\nu}= \frac12\left(T^{\mu\alpha\nu}+T^{\nu\mu\alpha}+T^{\alpha\mu\nu}\right)+g^{\alpha\mu}T^{\nu}-g^{\alpha\nu}T^{\mu}
\end{equation}
and
\begin{equation}
{\mathbb T}  =  \frac12 T_{\alpha\beta\mu}S^{\alpha\beta\mu}
 =  \frac14 T_{\alpha\beta\mu}T^{\alpha\beta\mu}+\frac12 T_{\alpha\beta\mu}T^{\beta\alpha\mu}-T_{\mu}T^{\mu}
\end{equation}
respectively, with the torsion vector being $T_{\mu}\equiv T^{\alpha}_{\hphantom{\alpha}\mu\alpha}$. Finally, $\Theta_{\mu\nu}$ is the energy-momentum tensor of the matter content of the universe (from now on assumed to be a perfect fluid).

We consider perturbations around the simplest cosmological solutions of the form
$$ds^2=a^2(\tau)\cdot\left(-d\tau^2+dx^i\cdot dx^i\right)$$
with the most trivial choice of the tetrad
$$e^A_{\mu}=a(\tau)\cdot\delta^A_{\mu}$$
and zero spin connection. Analysis in the Ref. \cite{GK} has been made in the pure tetrad gauge, i.e. $\omega^A_{\hphantom{A}\mu B}=0$ even for perturbations. (Note that due to the lack of local Lorentz invariance, the choice of this background tetrad in pure tetrad formalism is a part of the physical choice of background cosmology, not merely a possible description of a given solution; and it would be interesting to also construct perturbation theory around other solutions with the same metric, such as those of the Ref. \cite{BFG}.)

We used the following parametrisation for the tetrad variations
\begin{eqnarray*}
e^{\cancel{0}}_0 & = & a(\tau)\cdot\left(1+\phi\right),\\
e^{\cancel{0}}_i & = & a(\tau)\cdot\left( \partial_i \beta+u_i\right),\\
e^a_0 & = & a(\tau)\cdot\left( \partial_a \zeta+v_a\right),\\
e^a_j & = & a(\tau)\cdot \left((1-\psi)\delta^a_j+\partial^2_{aj}\sigma+\epsilon_{ajk}\partial_k s+\partial_j c_a+\epsilon_{ajk}w_k+\frac12 h_{aj}\right)
\end{eqnarray*}
with the usual assumptions that the vectors are divergenceless, and the tensor is divergenceless and traceless. The spacetime indices like $\mu$ run over $0$ for time and $i$ or $j$ for spatial coordinates, while for the tangent indices $A$ we use $\cancel{0}$ and $a$ respectively.

Apart from the perturbed metric components
\begin{eqnarray*}
g_{00} & = & -a^2(\tau)\cdot\left(1+2\phi\right),\\
g_{0i} & = & a^2(\tau)\cdot\left( \partial_i \left(\zeta-\beta\right)+v_i-u_i\right),\\
g_{ij} & = & a^2(\tau)\cdot \left((1-2\psi)\delta_{ij}+2\partial^2_{ij}\sigma++\partial_i c_j + \partial_j c_i +h_{ij}\right),
\end{eqnarray*}
we have 6 new variables related to local Lorentz transformations: boosts in 3 independent directions parametrised by $u_i+v_i+\partial_i ({\beta}+\zeta)$ and rotations around 3 independent axes parametrised by $w_i+\partial_i s$.

For the purposes of perturbation theory, it was convenient to separate the antisymmetric
\begin{equation}
\label{aseom}
 \left(S_{\nu\mu\alpha}-S_{\mu\nu\alpha}\right)\partial^{\alpha}\mathbb T=\left(T_{\alpha\mu\nu}+g_{\alpha\mu}T_{\nu}-g_{\alpha\nu}T_{\mu}\right)\partial^{\alpha}{\mathbb T}=0
\end{equation}
(in case $f_{TT}\neq 0$), and the symmetric 
\begin{equation}
\label{seom}
f_{T}\mathop{G^{\mu}_{\nu}}\limits^{(0)}+f_{TT}Q^{\mu}_{\nu}+\frac12 \left(f-f_{T}{\mathbb T}\right)\delta^{\mu}_{\nu}=8\pi G \Theta^{\mu}_{\nu}
\end{equation}
components of the equations of motion (\ref{eom}) where
\begin{equation}
\label{Q}
Q_{\mu\nu}\equiv\frac12 \left(S_{\mu\nu\alpha}+S_{\nu\mu\alpha}\right)\partial^{\alpha}\mathbb T.
\end{equation}

We refer the reader to the Ref. \cite{GK} for details (some of them will be reproduced in the Section 4 with inclusion of the spin connection). However the general result is that the linear perturbation theory is very similar to the case of general relativity with a notable exception of a gravitational slip $\phi-\psi\neq 0$ in absence of anisotropic stress. Here we would like to correct a number of small mistakes from that paper \cite{GK}.

\subsection{Pseudoscalar variation}
It was claimed in our previous work \cite{GK} that the pseudoscalar perturbation $s$ contributes to the variation of $T_{i0j}$ components
$$T_{i0j}=-T_{ij0}=a^2\cdot \left(H\delta_{ij}+\epsilon_{ijk}\partial_k s^{\prime}\right)$$
only ($H\equiv\frac{a^{\prime}}{a}$). It is not correct. One can easily derive that
$$T_{ijk}=-a^2\cdot (\epsilon_{ijl}\partial^2_{kl}s-\epsilon_{ikl}\partial^2_{jl}s)$$
which generically is not zero as can be seen by e.g. taking divergence $\partial_k T_{ijk}=-a^2\epsilon_{ijl}\partial_l \bigtriangleup s$.

However, one can see that the main claim was correct. The pseudoscalar perturbation does not show up in the linearised equations of motion. It is very easy to prove. The only place it can appear in is the perturbation of $S_{\nu\mu\alpha}$ in the  $f_{TT}S_{\nu\mu\alpha}\partial^{\alpha}{\mathbb T}$ term because the variation of the torsion scalar is easily computed to be $\delta{\mathbb T}=-\frac{4H}{a^2}\left(\bigtriangleup\zeta+3H\phi+3\psi^{\prime}\right)$ which does not contain $s$. And it is easily seen that $S_{\mu\nu 0}$ components do not depend on $s$ either.

This is actually an interesting point concerning the poorly understood issue of the so-called remnant symmetry \cite{FerrFio}. We see that out of the six new variables only one ($\zeta$) contributes to the linear variation of the torsion scalar. At the level of linearised equations of motion five of them contribute (all but $s$).

\subsection{The vector sector}
A more serious mistake has been made concerning the vector (and pseudovector) perturbations. We had the following (correct) torsion tensor components
\begin{eqnarray*}
T_{0ij} & = & a^2\cdot \left(\partial_j u_i-\partial_i u_j\right),\\
T_{00i} & = & a^2\cdot\left(-u_i^{\prime}+H(v_i-u_i)\right),\\
T_{ijk} & = & a^2\cdot \left(\epsilon_{ikl}\partial_j w_l-\epsilon_{ijl}\partial_k w_l\right),\\
T_{i0j} & = & a^2\cdot\left(H\delta_{ij}+\epsilon_{ijk}w_k^{\prime}-\partial_j v_i\right),
\end{eqnarray*}
with an obvious misprint in the form of unwanted factor of $\partial_i$ in $T_{00i}$ in Ref. \cite{GK}.
However, already the torsion vector was found incorrectly. Using the formulae above, we see
$$T_i=-u_i^{\prime}+\epsilon_{ijk}\partial_j w_k$$
which differs from the Ref. \cite{GK} by the first term, while it is of course correct that $T_0=3H$ does not receive any correction from the vector sector at linear order, and neither does the torsion scalar.

The antisymmetric equation of motion (\ref{aseom}) in its spatial part indeed boils down \cite{GK} to $T_{0ij}=0$ for which the only admissible solution in perturbation theory is $u_i=0$. However, the mixed components of this equation  (\ref{aseom}) were found incorrectly. Let us rewrite them as
\begin{equation}
\label{newvec}
0=\left(-T_{\alpha i}^{\hphantom{\alpha i}0}+\delta_{\alpha}^{0}T_{i}-g_{\alpha i}T_{0}\right)g^{\alpha 0}{\mathbb T}^{\prime}=(-T_{\hphantom{0} i}^{0 \hphantom{ i}0}+g^{00}T_i){\mathbb T}^{\prime}.
\end{equation}
One can easily see that $T_{\hphantom{0} i}^{0 \hphantom{ i}0}=\frac{1}{a^2}u_i^{\prime}$, and this equation gives $\epsilon_{ijk}\partial_j w_k=0$, to be solved as $w_i=0$, if  ${\mathbb T}^{\prime}\neq 0$. Given that
$${\mathbb T}=\frac{6H^2}{a^2},$$
the last condition is satisfied as long as the cosmology is different from Minkowski and de Sitter spacetimes where the model reduces to general relativity.

In the symmetric part of equations, the $Q$ tensor (\ref{Q}) is
$$Q_{\mu\nu}=\frac12 \left(\vphantom{\frac12}T_{\mu\nu\alpha}+T_{\nu\mu\alpha}+2g_{\mu\nu}T_{\alpha}-\left(g_{\alpha\mu}T_{\nu}+g_{\alpha\nu}T_{\mu}\right)\right)g^{\alpha 0}{\mathbb T}^{\prime}$$
And we find that the antisymmetric equation (\ref{newvec}) has set $Q^0_i$ to zero since
$$Q^0_i=\frac12 \left(T^{0\hphantom{i}0}_{\hphantom{0}i}-g^{00}T_i\right){\mathbb T}^{\prime}=\frac{6H(H^{\prime}-H^2)}{a^4}\epsilon_{ijk}\partial_j w_k.$$

Unbelievably enough, the $Q^0_i=0$ equality means that, according to mixed components of equation (\ref{seom}), despite all mistakes in calculations the final result given in the Ref. \cite{GK} for the constraint relating the metric perturbation and the vortical velocity $\mathfrak u$ of perfect fluid
$$f_{T}\bigtriangleup v_i=16\pi G a (\rho+p){\mathfrak u}_i$$
is absolutely correct. Even though $Q_{0i}\neq 0$, contrary to what has been stated there.

Calculations leading to the decay equation for vector perturbations are not influenced by these corrections, and it appears to be correct in the Ref. \cite{GK}. With this remark we conlcude the correction of mistakes.

\section{Variation of the spin connection}
Now we want to add to this game a variation of the spin connection in the flat class. It can be easily done. We just need to remember that at the linearised level around zero spin connection the flatness condition reads $\partial_{\mu}\omega^A_{\hphantom{A}\nu B}-\partial_{\nu}\omega^A_{\hphantom{A}\mu B}=0$ (quadratic terms omitted). As such, it is equivalent to flatness of six independent differential forms taking values in the field of real numbers. Since the assumed spacetime topology is trivial, the only solution is that the forms are exact: $\omega^A_{\hphantom{A}\nu B}=\partial_{\nu}\lambda^A_{\hphantom{A}B}$ with arbitrary $\lambda_{AB}=-\lambda_{BA}$. This is also the solution reported in the Ref. \cite{Topor}.

One can recognise the (linearised around $\Lambda=\mathbb I$) condition of the connection being inertial $\omega^A_{\hphantom{A}\nu B}={(\Lambda^{-1})}^A_C\partial_{\nu}\Lambda^C_B$, with $\lambda$ being an element of the Lie algebra of the Lie group to which $\Lambda$ belongs. In other words, flat connections are the same as inertial connections. In general, existence of flat non-inertial connections would be a cohomology problem for differential forms taking values in the Lorentz algebra. This freedom is connected to global issues, and therefore is expected to be of no big deal anyway. Moreover, these questions of global topology should not be considered separately from the global parallelisability condition of teleparallel approaches (not so restrictive in causal 3+1 dimensions though).

Taking these perturbations into accouint is a very trivial task. Indeed, by the very construction, our model, as well as any other covariant teleparallel model, is invariant under the following transformation:
\begin{equation}
\label{gaugeom}
e^A_\mu  \longrightarrow \Lambda^A_C e^C_\mu , \quad
\omega^A_{\hphantom{A}\mu B} \longrightarrow \Lambda^A_C \omega^C_{\hphantom{C}\mu D}(\Lambda^{-1})^D_B-(\Lambda^{-1})^A_C\partial_{\mu}\Lambda^C_B,
\end{equation}
which at the linear order around our background reads 
$$\delta e^A_{\mu}\longrightarrow \delta e^A_{\mu}+a(\tau)\cdot\lambda^A_{\hphantom{A}C}\delta^C_{\mu}, \quad \omega^A_{\hphantom{A}\nu B}\longrightarrow \omega^A_{\hphantom{A}\nu B}-\partial_{\nu}\lambda^A_{\hphantom{A}B}.$$
Therefore we see that one linear combination of $\partial\left(\frac{1}{a}\delta e\right)$ and $\omega$ is gauge invariant under local Lorentz transformations while another lacks any physical meaning, see also Ref. \cite{Tspec}. Covariantising the perturbation theory of Ref. \cite{GK} amounts to substituting the antisymmetric perturbation of the tetrad with its covariant version.

\section{Perturbation theory\\ in Lorentz-covariant formulation of f(T) gravity}
Let us illustrate the previous statements by explicit calculations. With $\omega^A_{\hphantom{A}\nu B}=\partial_{\nu}\lambda^A_{\hphantom{A}B}$ we see that very simple corrections have to be added to the torsion tensor components found in the Ref. \cite{GK}:
$$T_{\alpha\mu\nu}=\left.T_{\alpha\mu\nu}\vphantom{\int}\right|_{\omega=0}+a^2(\partial_{\mu}\lambda_{\alpha\nu}-\partial_{\nu}\lambda_{\alpha\mu})$$
where the Latin indices are traded for Greek ones in the usual way via applying the tetrad.

The spin connection fluctuations can be classified into the scalar and vector ones in precise analogy with the Lorentzian part of the tetrad:
\begin{eqnarray*}
\lambda_{0i} & = & \partial_i\tilde\zeta+\tilde u_i,\\
\lambda_{ij} & = & \epsilon_{ijk}(\partial_k \tilde s +\tilde w_k)
\end{eqnarray*}
 with divergenceless vectors $\tilde u$ and $\tilde w$. 

Note that it is very important to separate the scalar and the vector perturbations. It  has not been done in the Ref. \cite{Topor}, and that's how the Authors were led to their first attempt of solving the equations with a zero gravitational slip ansatz. They arrive at a condition that a curl equals a gradient. Since it requires in perturbation theory that both are zero, the system ends up being overdetermined in precise analogy with the old Ref. \cite{oldpert}. In both cases the antisymmetric perturbation is unjustifiably neglected, either by choosing the diagonal tetrad ansatz in the pure tetrad gauge, or by setting some spin connection components to zero at will in a gauge in which all antisymmetric perturbations are moved to the spin connection sector.

\subsection{Tensor perturbations}

Since there is no tensor perturbation in the spin connection, this equation remains the same:
$$f_Th_{ij}^{\prime\prime}+2H\left(f_T + \frac{6f_{TT}(H^{\prime}-H^2)}{a^2}\right)h_{ij}^{\prime}-f_T \bigtriangleup h_{ij}=0.$$

\subsection{Vector perturbations}
For vector perturbations with $c=0$ (diffeomorphisms') gauge, we get
\begin{eqnarray*}
T_{0ij} & = & a^2\cdot\left(\partial_j (u_i-\tilde u_i)-\partial_i (u_j-\tilde u_j)\right),\\
T_{00i} & = & a^2\cdot\left(-(u_i-\tilde u_i)^{\prime}+H(v_i-u_i)\right),\\
T_{ijk} & = & a^2\cdot \left(\epsilon_{ikl}\partial_j (w_l+\tilde w_l)-\epsilon_{ijl}\partial_k (w_l+\tilde w_l)\right),\\
T_{i0j} & = & a^2\cdot\left(H\delta_{ij}+\epsilon_{ijk}(w_k+\tilde w_k)^{\prime}-\partial_j (v_i-\tilde u_i)\right).
\end{eqnarray*}

All calculations proceed as before introducing the spin connection, with only difference that $w$ is replaced by $w+\tilde w$ (so that we can find this sum of variables but not their difference which is free due to local rotational symmetry in the covariant formulation), and analogously $v+u$ is replaced by $v+u-2\tilde u$, while the metric perturbation $v-u$ is intact. The antisymmetric equations give now $u_i-\tilde u_i=0$ and $w_i+\tilde w_i=0$.

Since the same combinations, $w+\tilde w$, $v+u-2\tilde u$ and $v-u$, enter the symmetric part of equations, the latter finally experiences no change:
$$f_{T}\bigtriangleup (v_i-u_i)=16\pi G a (\rho+p){\mathfrak u}_i$$
and 
$$f_{T}\cdot (v_i-u_i)^{\prime}+2\left(f_T H +\frac{6f_{TT}H(H^{\prime}-H^2)}{a^2}\right)\cdot(v_i-u_i)=0$$
for the decay equation. The only subtlety is that now we have to write the genuine metric perturbation $v-u$ instead of simple $v$ because the $u=0$ equality (instead of $u=\tilde u$) is just a gauge choice of vanishing spin connection.

\subsection{Pseudoscalar perturbations}
In pseudoscalars we have
\begin{eqnarray*}
T_{i0j} &  = & a^2\cdot\left(H\delta_{ij}+\epsilon_{ijk}\partial_k (s+\tilde s)^{\prime}\right),\\
T_{ijk} & = & -a^2\cdot(\epsilon_{ijl}\partial^2_{kl}(s+\tilde s)-\epsilon_{ikl}\partial^2_{jl}(s+\tilde s)).
\end{eqnarray*}

Of course, they still do not make it to the linearised field equations. Both $s$ and $\tilde s$ remain undetermined, the difference due to the rotational symmetry in covariantised models, and the sum because of a mysterious reason which can be called remnant symmetry. Technically the latter can be traced back to the fact that $S_{ij0}$ accidentally appears identically symmetric at the linear order, so that we lack one equation in the antisymmetric part.

\subsection{Scalar perturbations}

Finally, for scalar perturbations, in the $\sigma=0$ and $\beta=\zeta$ (diffeomorphisms') Newtonian gauge, we have
\begin{eqnarray*}
T_{00i} & = & a^2\cdot \partial_i\left(\phi-(\zeta-\tilde\zeta)^{\prime}\right),\\
T_{ijk} & = & a^2\cdot \left(\delta_{ij}\partial_k \psi-\delta_{ik}\partial_j\psi\right),\\
T_{i0j} & = & a^2\cdot\left(H\delta_{ij}- \partial^2_{ij}(\zeta-\tilde\zeta)- \delta_{ij}\left(2H\psi+\psi^{\prime}\right)\right),
\end{eqnarray*}
so that $\zeta$ of the Ref. \cite{GK} is replaced by $\zeta-\tilde\zeta$.

Again, the meaning is simple. The sum $\zeta+\tilde\zeta$ is a pure (Lorentz group) gauge variable, while the difference can be found via the antisymmetric equation:
\begin{equation}
\label{zeta}
\bigtriangleup(\zeta-\tilde\zeta)=-3\left(\psi^{\prime}+H\phi-\frac{H^{\prime}-H^2}{H}\psi\right).
\end{equation}
And this very same combination enters all the equations of the symmetric part. For example, the off-diagonal one gives the gravitational slip as:
\begin{equation}
\label{slip}
\phi-\psi=-\frac{12f_{TT}H(H^{\prime} - H^2)}{f_{T}}(\zeta-\tilde\zeta).
\end{equation}
Obviously, after eliminating the $\zeta$-s, neither this nor any other scalar perturbation equation of the Ref. \cite{GK} experiences any modification.

Let us remark again that including the spin connection into the cosmological perturbation theory amounts to rewriting the very same equations of the pure tetrad formalism in a (local Lorentz) covariant form.  From this point of view, the approach of Refs. \cite{GK, BarSot} corresponds to the $\tilde\zeta=0$ gauge in the scalar sector, while the Ref. \cite{Topor} employs the $\zeta=0$ gauge instead. If correctly done, the latter method can lead to nothing but the very same results as the pure tetrad approach did.

An important point however is that, one way or another, the antisymmetric part of the perturbation must be fully taken into account. Otherwise calculations would not be self-consistent, and the equation of motion (\ref{eom}) might turn out overdetermined \cite{oldpert}.

\section{Conclusions}

We have shown by explicit calculations that the perturbation theory in $f({\mathbb T})$ cosmology does not depend on whether one works in the pure tetrad formalism or in a covariant formulation. Of course, this is not a big surprise. However, it allowed us to present the way of working with cosmological perturbations in presence of the flat spin connection. And at the same time we corrected some small mistakes from the Ref. \cite{GK}. Fortunately, they did not change the main conclusions, and the worst ones belonged to the least interesting sector of perturbations, the vector one. 

Note that the approach presented here is completely general.  Introducing the flat spin conneciton leads to the same (local Lorentz) gauge invariant combinations of tetrad and spin connection components \cite{Tspec} in perturbation theory of any modified teleparallel model \cite{GK, Rocco}, if properly covariantised \cite{GKS} indeed. These considerations should be taken seriously if one wishes to study cosmological implications of teleparallel approaches to modified gravity.

{\bf Acknowledgements.} I am grateful to M.J. Guzm\'an and T.S. Koivisto for their comments on the initial draft of this manuscript.

\end{document}